\documentclass[aps,pre]{revtex4}
\usepackage{epsfig}
\usepackage{amsmath}
\usepackage{amssymb}
\begin{document}

\title{\bf\noindent The role of the interaction matrix in mean-field
spin glass models} 

\author{R. Cherrier, D.S. Dean and A. Lef\`evre}
\affiliation{IRSAMC, Laboratoire de Physique Quantique, Universit\'e Paul
Sabatier, 118 route de Narbonne, 31062 Toulouse Cedex 04, France.}

\begin{abstract}
Mean-field models of 2-spin Ising spin glasses with interaction matrices
taken from ensembles which are invariant under $O(N)$ transformations
are studied. A general study shows that the nature of the spin glass 
transition can be deduced from the eigenvalue spectrum of the 
interaction matrix. A simple replica approach is derived to carry out 
the average over the $O(N)$ disorder.
The analytic results are confirmed by extensive Monte Carlo simulations
for large system sizes and by exact enumeration for small system sizes.  
\end{abstract}
\date{26 November 2002}
\maketitle

\section{Introduction}
Mean-field models of spin glasses have been extensively studied
over the last 30 years \cite{review}. The first mean-field model to be studied
thoroughly was the Sherrington-Kirkpatrick \cite{sk} model which exhibits
a classical spin glass transition with a continuous transition in the 
Parisi overlap matrix $Q_{ab}$ at the transition temperature $T_c$. 
The full solution to this problem requires continuous replica symmetry 
breaking \cite{parisi}, indicating 
an extensive number of pure states in the low temperature phase. 
Mean-field models with multi or $p$-spin, interactions exhibit 
discontinuous jumps in the Parisi overlap matrix $Q_{ab}$ at the static
transition temperature denoted by $T_K$ 
for $p>2$ \cite{crisom}. However these systems exhibit a dynamical
transition at a temperature $T_D > T_K$ indicating the onset of
an extensive number of metastable states preceding the static transition. 
These
models are of particular interest as the scenario of a dynamical transition
followed by a static transition is observed in structural glasses
\cite{kt,ktw}. For this reason the above type of behavior is often referred 
to as a structural glass transition. Potts type
spin glasses can also exhibit first order phase transitions
\cite{review}. In this paper
we restrict ourselves to the study of spin glasses with 2-spin interactions
and concentrate on the role of the interaction matrix in determining
the nature of the phase transitions in the system. Mean-field spin glass
type models appear in a wide range of contexts, they are of course
the starting points for studying models of finite dimensional spin glasses but
also arise as  models of  neural networks, formulations
of optimization problems and simple models for protein folding. 

We shall analyze a class of mean field spin glass
models with Hamiltonian 
\begin{equation}
H = -\frac{1}{2}\sum_{ij} J_{ij} S_i S_j \label{H}
\end{equation}
where the $S_i$ are $N$ Ising spins.
The interaction matrix $J$ is constructed via the following
procedure
\begin{equation}
J = {\cal O}^T\Lambda  {\cal O} \label{J}
\end{equation}
where ${\cal O}$ is a random $O(N)$ matrix chosen with the Haar
measure. The matrix $\Lambda$ is diagonal with elements
independently chosen from a distribution $\rho(\lambda)$. The 
support of $\rho(\lambda)$ is taken to be finite and independent of $N$, this
ensures the existence of the thermodynamic limit. 
The interest of this kind
of model is that one may average over the $O(N)$ disorder ${\cal O}$ and
then examine the nature of the spin glass phase as a function of the 
eigenvalue distribution $\rho(\lambda)$. In particular we shall 
show that the way in which  $\rho(\lambda)$ vanishes at the maximal
value of its support, $\lambda_{\max}$, determines whether the glass transition
is a classical spin glass transition or a structural glass transition.
We show that a finite temperature classical spin glass transition 
occurs if the same 
model but with spherically constrained spins (such that 
$S_i \in (-\infty,\infty)$ and $\sum_i S_i^2 = N$) 
exhibits a finite temperature phase transition. Where this 
is not the case we study the system using  a one step replica symmetry breaking
scheme to determine the dynamical transition temperature $T_D$  and 
the Kauzmann temperature $T_K$. Numerical 
simulations are carried out to confirm our analytical predictions on this
class of models. We carry out both Monte Carlo simulations and
exact enumeration calculations. The dynamical
transition temperature $T_D$ estimated from simulations agrees well
with our analytic calculations. The exact enumeration carried out
on small system sizes confirms the dynamical nature of the transition 
occurring at $T_D$. 

Let us  briefly recall some well studied models which fall into the 
class of spin glass models with interaction matrix given by the form
of Eq. (\ref{J}).
The Sherrington-Kirkpatrick \cite{sk} (SK) model with $J$ 
taken from the Gaussian
ensemble $J_{ij} = J_{ji}$ and  $J_{ij}$ independent Gaussian 
random variables  of zero mean and
with $\overline{J_{ij}^2} = 1/N$ can also  can be written in the form 
of Eq. (\ref{J}) with the Wigner
semi-circle law \cite{mehta}  density of eigenvalues given by
\begin{equation}
\rho(\lambda) = \frac{\left(4 - \lambda^2\right)^\frac{1}{2}}{2\pi}
\label{dossk}
\end{equation}
The Squared Interaction Matrix SK (SIMSK) model studied recently in \cite{deri}
has interaction matrix $J' = J^{T}J$, where the interaction matrix $J$
is taken from the Gaussian ensemble described above. Here $J'$ is also of the 
form given by Eq. (\ref{J}). The density of eigenvalues here is given by
\begin{equation}
\rho(\lambda) = {\left(4 - \lambda\right)^\frac{1}{2}
\over 2\pi\lambda^{1\over2}}\label{dossim}
\end{equation}
In fact the SIMSK model, at positive temperatures, is equivalent to the 
Hopfield model \cite{hop} with $N$ patterns. This model was 
shown \cite{deri} to have
different behavior at positive and negative temperatures. 
In the positive temperature
Hopfield  model \cite{amit}, the transition is a classical  spin glass 
transition as in the
SK model. However at negative temperature the model has a
structural glass transition \cite{deri}. 
We also note that the minority game, which is an economics model,
is closely related to the negative temperature or antiferromagnetic
Hopfield model and the same structural glass transition has been remarked \cite{dm}.     
In both the SK and SIMSK models
one knows that the eigenvalues of $J$ (the diagonal 
elements of $\Lambda$) are correlated
\cite{mehta}, however we will  see here that
in the thermodynamic limit this correlation seems to be unimportant.
One may also consider the more general Hopfield model with interaction matrix
\begin{equation}
J_{ij} = \sum_{\mu = 1}^{p} x^\mu_i x^\mu_j
\end{equation} 
where  $p = \alpha N$, for $\alpha$ of order $1$, is the number of patterns.
The case where $x_i^{\mu}$ are Gaussian random variables of zero mean with
correlation $\overline{ x^\mu_i x^\nu_j} = \delta_{ij}\delta^{\mu\nu}/N$
also falls into the class of models we are considering, as an arbitrary
orthogonal transformation ${\bf x}^{\mu} \to {\cal{O}}{\bf x}^{\mu}$
gives an element in the same statistical ensemble. Here the 
density of eigenvalues of the matrix $J$ is \cite{rhohopf} 
\begin{eqnarray}
\rho(\lambda) &=& {\left( 4\lambda - (\lambda +1 -\alpha)^2\right)^{1\over2}
\over 2\pi\lambda} + (1-\alpha)\delta(\lambda)\ \ {\rm for} \ \alpha < 1\ \;
\lambda = 0 \ {\rm and}\ \lambda \in [(1-\sqrt{\alpha})^2,(1+\sqrt{\alpha})^2] 
\label{doshl1}\\
&=& {\left( 4\lambda - (\lambda +1 -\alpha)^2\right)^{1\over2}
\over 2\pi\lambda} \ \ {\rm for} \ \alpha \ge 1 
\ ;\lambda \in [(\sqrt{\alpha}-1)^2,(1+\sqrt{\alpha})^2]
\label{doshg1}
\end{eqnarray}
Hence in the case $\alpha > 1$ the density of eigenvalues is zero at the 
extremes of the support of $\rho(\lambda)$. In the case  $\alpha \leq 1$,
the density of eigenvalues is non zero, and in fact diverges, at the lower
band edge but stays zero at the upper band edge. We remark that  
the density of eigenvalues Eq. (\ref{doshg1}) when $\alpha = 1$ is
exactly the same density of eigenvalues as in the SIMSK model, as
expected from our earlier discussion.

Another example is the Random Orthogonal Model (ROM) studied by Marinari,
Parisi and Ritort \cite{rom}, where 
\begin{equation}
\rho(\lambda) = \alpha \delta(\lambda-1) + (1-\alpha) \delta(\lambda+1)
\label{dosrom}
\end{equation}
This model  was extensively investigated in the case $\alpha = 1/2$,
and was shown to exhibit a structural glass transition. The case
$\alpha = 1/2$ is of particular interest because the high temperature
series expansion in this case is equivalent to that of 
a frustrated mean-field model, 
the sine model, which has no quenched
disorder. The ROM at $\alpha = 1/2$ shows some rather interesting behavior,
the static transition temperature $T_K$ is 
extremely close to the temperature $T_A$ where the annealed entropy vanishes. 
Below the static transition temperature the energy is almost constant 
or equivalently the specific heat is nearly zero. This implies
that the ROM at $\alpha = 1/2$ is {\em almost} a Random Energy Model
(REM) at the static level. The simplest version of the 
REM \cite{derrida} is given by considering
a system with micro-states having independent energies. This situation
arises by construction in the REM of Derrida where there are $2^N$ micro-states
$\nu$ of energies $E_\nu$ 
each chosen independently from a suitable distribution.
The REM also arises when one considers the $p\to \infty$ limit of $p$-spin
interaction mean field spin glasses \cite{derrida,gross}. Another example is
the case of directed polymers on Cayley trees
with random bond or site disorder \cite{dpct}. Although in the 
directed polymer problem there are 
correlations between paths, these correlations are weak and the
resulting thermodynamics is also REM-like. In this paper we will
show that this REM like behavior is enhanced in the ROM model on increasing
$\alpha$ above $\alpha = 1/2$.

At a more technical level, 
the problem of averaging over the $O(N)$ disorder was solved by
Marinari et al \cite{rom} by transposing results of Itzykson and Zuber 
\cite{itzu} (based on generating function techniques), 
from random matrix theory. For completeness we also give a simple physical
(though not rigorous) re-derivation of these averaging results.
 
\section{Averaging over the disorder}

We consider the partition function for a model with Gaussian
spins  with a random  interaction matrix $J$ with
density of eigenvalues $\lambda$ denoted by $\rho(\lambda)$. The 
partition function at $\beta = 1$ is given by

\begin{equation}
Z = \int \prod_i \ dS_i \exp\left({1\over 2} \sum_{ij}J_{ij} S_i S_j
-{\mu \over 2} \sum_i S_i^2 \right)
\end{equation}
The partition function may be explicitly evaluated, as in the case of
the $p=2$ spherical spin glass model \cite{p2}, by passing to the basis
of eigenvalues of the matrix $J$ 
\begin{equation}
Z = \int \prod_{\lambda} dS_{\lambda} 
\exp\left( {1\over 2} \sum_{\lambda} \lambda S_{\lambda}^2 -
{\mu\over 2}\sum_{\lambda} S_\lambda^2\right)
\end{equation}
The Gaussian integrals are easily performed yielding
\begin{equation}
Z = (2\pi)^{{N\over 2}} \prod_{\lambda} {1\over (\mu -\lambda)^{1\over 2}}
\end{equation}
thus 
\begin{equation}
\ln(Z) = {N\over 2} \ln(2\pi) -{1\over 2} \sum_{\lambda} \ln(\mu -\lambda)
\end{equation}
Averaging over the disorder we obtain
\begin{equation}
g = \overline{{\ln(Z)\over N}} = {1\over 2}\ln(2 \pi) -{1\over 2}
\int d\lambda \rho(\lambda) \ln(\mu -\lambda)\label{exact}
\end{equation}
We will now repeat the same calculation of $g$ using the replica method. One
replicates the system $n$ times, where we shall consider the limit
$n\to 0$:
\begin{equation}
Z^n = \int \prod_{i,a} dS_i^a \exp\left( {1\over 2}
\sum_{ij} J_{ij} \sum_{a} S_i^a S_i^a - {\mu\over 2}\sum_{i,a} {S_i^a}^2
\right)
\end{equation}
where $a= 1,\cdots, n$ are replica indices.
In a model where the interaction matrix is chosen to give an extensive free
energy, we expect that
\begin{equation}
\overline{\exp\left({1\over 2}\sum_{ij} J_{ij} \sum_a^n S_i^a S_j^a\right)}
= \exp\left({N\over 2}{\rm Tr} G(Q) + {\rm n.e.t.}\right)
\end{equation}
where Tr indicates the matricial trace over the Parisi order parameter 
matrix $Q_{ab} = {1\over N} \sum_i S_i^a S_i^b$, and the term  n.e.t.
denotes non extensive terms. The idea of the calculation that follows 
is to calculate $g$ using the replica method and then extract $G$ by comparing
the result of this replica calculation with the result (\ref{exact}). 

One has therefore for a generic $G$
\begin{equation}
\overline{Z^n} \sim \int \prod_{i,a} dS_i^a
\exp\left( {N\over 2} {\rm Tr} G(Q)  -{\mu\over 2} \sum_{i,a}{S_i^a}^2
\right)
\end{equation}
We impose the constraint $ NQ_{ab} = \sum_i S_i^a S_i^b$ with a 
Fourier representation of the delta function to obtain
\begin{eqnarray}
\overline{Z^n}  &\sim& \int \prod_{a,b} d\Lambda_{ab} dQ_{ab}
\prod_{i,n} dS_i^a \exp\left( {N\over 2} {\rm Tr} G(Q) + {N\over 2}
{\rm Tr} \Lambda Q -{1\over 2} \sum_{ab} \Lambda_{ab} \sum_{i} S_i^a S_i^b
-{\mu \over 2} \sum_{i,a} {S_i^a}^2 \right) \nonumber \\
&\sim& \int \prod_{a,b} d\Lambda_{ab} dQ_{ab} 
\exp\left(N S^*(Q,\Lambda)\right)
\end{eqnarray}
where the action $S^*(Q,\Lambda)$ over the order parameters $Q$ and 
$\Lambda$ is given by
\begin{equation}
S^*(Q,\Lambda) = {1\over 2} \left[ {\rm Tr} G(Q) + {\rm Tr} Q\Lambda
- {\rm Tr} \ln\left( \Lambda + \mu I\right)  + n\ln(2\pi)\right]
\end{equation} 
The saddle point equations $\partial S^* /\partial \Lambda_{ab} = 0$
yield the relation $Q = (\Lambda + \mu I)^{-1}$  thus giving the 
result
\begin{equation}
{\ln\left(\overline{Z^n}\right)\over nN}
= {1\over 2}\ln(2\pi) + {1\over 2} -{1\over 2 n} {\rm extr}_{Q}
\left[  \mu {\rm Tr} Q -{\rm Tr} G(Q) - {\rm Tr} \ln(Q) \right]
\label{act1}
\end{equation}
where $\rm{extr}_Q$ indicates that the function in the square brackets
is evaluated at an extremal or stationary point. For integer $n$ 
this extremal value is of course the maximum, however in the limit $n\to 0$
it is often the minimal value that should be taken. The nature of the
stationary point chosen depends on the stability analysis of the Hessian
matrix at that point.

We now consider what form of ansatz one should make for $Q$ in the
variational problem contained in Eq. (\ref{act1}). The physical nature
of the problem makes it clear that the ansatz should be replica symmetric,
the system minimizes its energy on condensing near the maximal 
eigenvalues of the matrix $J$ and there is no frustration. We make
the ansatz $Q = q_0 I + q U$ where $U_{ab} =1$ for all $a$, $b$.
Making use of the fact that $U^2 = n U$, in the limit $n\to 0$ we
obtain
\begin{eqnarray}
g &=& \lim_{n\to 0} {\ln(\overline{Z^n}) \over nN}
\nonumber \\
&=& {1\over 2} + {1\over 2}\ln(2\pi) - {1\over 2} {\rm extr}_{q_0,q}
[  \mu(q_0 +q) - G(q_0) - q G'(q_0)- \ln(q_0) - {q\over q_0} ]
\end{eqnarray}
The stationarity condition with respect to $q$ yields 
$G'(q_0) - \mu + {1\over q_0} = 0 $ which then gives
\begin{equation}
g = {1\over 2} + {1\over 2}\ln(2\pi) - {1\over 2} {\rm extr}_{q_0}
[ \mu q_0 -G(q_0) - \ln(q_0) ] \label{g2}
\end{equation}
If one returns to expression for the action in Eq. (\ref{act1}), it is
easy to understand the result in Eq. (\ref{g2}). The term in square brackets
in 
Eq. (\ref{act1}) clearly possesses an $O(n)$ invariance which is 
a consequence of the $O(N)$ invariance of the original problem 
before the disorder average is carried out. The action  (\ref{act1}) can 
therefore be written in terms of the eigenvalues of the matrix of 
$Q_{ab}$, which by comparison with Eq. (\ref{g2}) must correspond to
the possible values of $q_0$.  
We now equate the two different calculations for $g$, Eq. (\ref{exact})
and Eq. (\ref{g2}) to obtain
\begin{equation}
{\rm min}_{q_0} [ \mu q_0 -G(q_0) - \ln(q_0) ]
= 1 + \int d\lambda \rho(\lambda) \ln(\mu - \lambda) \label{leg}
\end{equation}
The function 
\begin{equation}
f(\mu) = 1 + \int d\lambda \rho(\lambda) \ln(\mu - \lambda) 
\end{equation}
is clearly concave for $\mu > \lambda_{\max}$, where $\lambda_{\max}$
is the largest eigenvalue of the interaction matrix $J$.
Hence the right hand side of Eq. (\ref{leg}) has the form 
of a Legendre transform which can now be inverted to give the result
\begin{equation}
G(z) = {\rm extr}_{\mu} [ \mu z -  \int d\lambda \rho(\lambda) \ln(\mu - \lambda)
] - \ln(z) -1 \label{eqGz}
\end{equation}
or explicitly
\begin{equation}
G(z) = z \mu(z) - \int d\lambda \rho(\lambda) \ln(\mu(z) - \lambda)
-\ln(z) - 1
\end{equation}
where $\mu(z)$ is given by the solution to
\begin{equation}
z = \int {\rho(\lambda) d\lambda \over \mu(z) -\lambda} \label{eqmu}
\end{equation}
The concavity of $f(\mu)$ furthermore assures the uniqueness of 
$\mu(z)$ and hence the annealed calculation (with $n=1$ replicas)
is equivalent to the quenched calculation (with $n=0$ replicas).
Hence the extremum taken in Eq. (\ref{eqGz}), should be a minimum.
Consequently we obtain the final result,
\begin{equation}
G(z) = {\rm min}_{\mu} [ \mu z -  \int d\lambda \rho(\lambda) \ln(\mu - \lambda)
] - \ln(z) -1  \label{ourg}
\end{equation}
This result can be shown to be identical to that used by Marinari et al
\cite{rom} who transposed the results of Itzykson and Zuber \cite{itzu}
for integrals over unitary matrices to integrals over orthogonal matrices.
We recall briefly the prescription of \cite{rom} in the form
adapted to the definition of the Hamiltonian used here
(there is a difference of definition by  a factor of $2$). 
In the method of \cite{rom} $G(z)$ is given
by
\begin{equation}
G(z) = \int_0^1 {\left( \psi(tz) -1 \right)\over t}\ dt \label{gma}
\end{equation}
where 
\begin{equation}
 \psi(z) = \int d\lambda \rho(\lambda) {1\over 1 - j(z) \lambda}\label{eqpsi}
\end{equation}
with $j(z)$ given by the solution to the equation
\begin{equation}
z = j(z) \int d\lambda \rho(\lambda) {1\over 1 - j(z) \lambda}
\label{eqjz}
\end{equation}
Comparison of Eq. (\ref{eqjz}) with Eq. (\ref{eqmu}) shows that 
$\mu(z) = 1/j(z)$. In addition one sees from Eqs. (\ref{eqpsi}) and 
(\ref{eqjz}) that $\psi(z) = z/j(z) = z\mu(z)$. When $z \ll 1$ one has the 
solution $\mu(z) \approx 1/z$, or $j(z) \approx  z$ from Eq. (\ref{eqmu}).
In both prescriptions this yields (as it should), $G(0) = 0$. One also
has that $\psi(0) = 1$, thus differentiating Eq. (\ref{gma}) yields
\begin{equation}
G'(z) = \mu(z) -{1\over z} \label{gprime}
\end{equation}
which is the same equation as obtained on differentiating our result
Eq. (\ref{ourg}). The equivalence of the two averaging results is thus
demonstrated. One of the advantages with the derivation of the averaging
formula derived  here is that it has a variational form.

Here we shall give some specific examples of $G(z)$ for some well 
known models and others we will study in this paper.
\begin{itemize}
\item {\bf The Sherrington-Kirkpatrick model:}
The first example to consider is the Sherrington-Kirkpatrick model, for
which the function G is known by simply averaging over the independent
Gaussian elements
of $J$: $G(z)=z^2/2$. We shall show how to get this result from 
the formalism developed above.

From Eq. (\ref{eqmu}):
\begin{eqnarray}
z &=& \frac{1}{2\pi} \int_{-2}^2 d\lambda
\frac{\sqrt{4-\lambda^2}}{\mu(z)-\lambda}\\ \nonumber 
 &=& \frac{\mu(z)-\sqrt{{\mu(z)}^2-4}}{2}.
\end{eqnarray}
Solving this gives
\begin{equation}
\mu(z)=z+\frac{1}{z}
\end{equation}
which gives $G(z)= z^2/2$ by using Eq. (\ref{gprime}).

\item {\bf The Hopfield model:}
For all $\alpha$  Eq. (\ref{eqmu}) yields
\begin{equation}
z = {1\over 2 \mu}\left[ \mu - \alpha + 1 - \left((\mu - \alpha + 1)^2 - 4 \mu
\right)^{1\over 2}\right] 
\end{equation}
The solution to this equation turns out to be surprisingly simple and is 
\begin{equation}
\mu = - {\alpha \over 1-z} + {1\over z}
\end{equation}
Again integrating Eq.(\ref{gprime}) gives
\begin{equation}
G(z) = -\alpha \ln(1-z)
\end{equation}
\item {\bf The ROM:} In the ROM  Eq. (\ref{eqmu}) reads

\begin{equation}
z = {\alpha\over \mu -1} + {1-\alpha\over \mu +1}
\end{equation}
Solving this yields
\begin{equation}
\mu = {1 \pm \left(1 + 4 z (m + z)\right)^{1\over 2}\over 2z }
\end{equation}
where $m = 2\alpha -1$.
The solution of $\mu$ should always be such that $\mu > \lambda_{max}$ hence
we take the positive root in the above equation.
The subsequent integration of  Eq.(\ref{gprime}) then gives

\begin{eqnarray}
G(z) &=& {1\over2} \left[\left(1 + 4 z (m + z)\right)^{1\over 2} + m
\ln\left(\left(1 + 4 z (m + z)\right)^{1\over 2} + 2 z + m\right)
\nonumber \right.\\
&-&\left.\ln\left(\left(1 + 4 z (m + z)\right)^{1\over 2} + 1 + 2 mz\right)
- m \ln(m + 1) -1 - \ln(2)\right]
\end{eqnarray}
Setting $m=0$ yields the symmetric case $\alpha =1/2$ \cite{rom}. For
this special case, the partition function may be computed directly by using
the $O(N)$ invariance \cite{brezin}.  

\item {\bf The semi-square law:}
The semi-square model is one with eigenvalues distributed uniformly between
$-1$ and $1$ and hence $\rho(\lambda) = 1/2$ for $\lambda \in [-1,1]$. In this
case the Eq. (\ref{eqmu}) is
\begin{equation}
z = {1\over 2}\ln\left({\mu + 1\over \mu -1}\right)
\end{equation}
This leads to 
\begin{equation}
G(z) = \ln\left({\sinh(z)\over z}\right)
\end{equation}
\end{itemize}

\section{The General Case}
\subsection{Representations of the Saddle Point Action}
We repeat the precedent calculation for an Ising spin
Hamiltonian of form given in
Eq. (\ref{H}). Using the same technique as
the previous section, after a little algebra, one finds
\begin{equation}
{\ln\left(\overline{Z^n}\right)\over N} = {\rm extr}_{Q,\Lambda} 
S^{**}[Q,\Lambda]
\end{equation}
where 
\begin{equation}
S^{**}[Q, \Lambda] = {1\over 2}{\rm Tr} G(\beta Q) + {1\over 2}{\rm Tr}Q\Lambda
+ \ln\left[ {\rm Tr}_{S_a} \exp\left(-{1\over 2}\sum_{a,b}
\Lambda_{ab}S_a S_b\right)\right] 
\end{equation}
this is the general form used in \cite{rom}. 
However, as the form of $G$ is in 
general rather complicated, one may use the variational representation
of Eq. (\ref{ourg}), introducing an additional order parameter matrix $R$
to write
\begin{equation}
{\ln\left(\overline{Z^n}\right)\over N} = {\rm extr}_{Q,\Lambda,R} 
S^{*}[Q,\Lambda, R]
\end{equation}
where
\begin{eqnarray}
S^{*}[Q, \Lambda,R] &=&   {\beta\over 2}{\rm Tr} QR -  {1\over 2} {\rm Tr}
\int d\lambda \rho(\lambda) \ln(R - \lambda)
 - {1\over 2}{\rm Tr}\ln(\beta Q) -{n\over 2} \nonumber \\
&+& {1\over 2}{\rm Tr}Q\Lambda
+ \ln\left[ {\rm Tr}_{S_a} \exp\left(-{1\over 2}\sum_{a,b}
\Lambda_{ab}S_a S_b\right)\right]
\end{eqnarray} 
The saddle point equation $\partial S^*/\partial Q = 0$ yields the relation
$\Lambda = Q^{-1} - \beta R$, this leads to 
\begin{equation}
{\ln\left(\overline{Z^n}\right)\over N} = {\rm extr}_{Q,R} 
S[Q, R]
\end{equation}
where 
\begin{equation}
S[Q, R] = -  {1\over 2} {\rm Tr}
\int d\lambda \rho(\lambda) \ln(R - \lambda) - {1\over 2}{\rm Tr}\ln(\beta Q) +  \ln\left[ {\rm Tr}_{S_a} 
\exp\left({1\over 2}\sum_{a,b}
\left( \beta R_{ab} - [Q]^{-1}_{ab}\right) 
S_a S_b\right)\right] \label{actr}
\end{equation}
The saddle point equations for this action yield
\begin{equation}
Q_{cd} ={ {\rm Tr}_{S_a} S_c S_d
\exp\left({1\over 2}\sum_{a,b}
\left( \beta R_{ab} - [Q]^{-1}_{ab}\right) 
S_a S_b\right) \over {\rm Tr}_{S_a}
\exp\left({1\over 2}\sum_{a,b}
\left( \beta R_{ab} - [Q]^{-1}_{ab}\right) 
S_a S_b\right)}
\end{equation}
and 
\begin{equation}
\beta Q = \int d\lambda \rho(\lambda) (R-\lambda)^{-1}
\end{equation}

The problem may also be formulated purely in terms of the Parisi overlap
matrix $Q$. In this version one has
\begin{equation}
{\ln\left(\overline{Z^n}\right)\over N} = {\rm extr}_{Q} 
S[Q]
\end{equation}
where 
\begin{equation}
S[Q]  = {1\over 2}{\rm Tr} G(\beta Q) - {\beta \over 2}{\rm Tr}QG'(\beta Q)
+ \ln\left[ {\rm Tr}_{S_a} \exp\left({\beta\over 2}\sum_{a,b}
[G'(\beta Q)]_{ab}S_a S_b\right)\right] \label{formgen}
\end{equation}
Here we can show that if one uses the density of eigenvalues
for the SK, SIMSK or Hopfield model in the above formula Eq. (\ref{formgen}),
the saddle point action for the corresponding model is reproduced. Any
effects due to correlations between eigenvalues presumably only show
up as finite size corrections.
\subsection{The Replica Symmetric and Annealed Cases}

We start by computing the annealed free energy which presumably is the correct
free energy at sufficiently high temperatures.  
In the annealed case, that is $n=1$, the free energy is given by:
\begin{equation}
f_{ann}=-\frac{\ln(2)}{\beta}-\frac{1}{2\beta} G(\beta)
\end{equation}
and the entropy:
\begin{equation}
s_{ann}=\ln(2)+\frac{1}{2}G(\beta)-\frac{\beta}{2}G'(\beta)
\end{equation}
In the replica symmetric (RS) ansatz $Q_{ab}=(1-q)\delta_{ab}+q$, where $\delta$
is the Kronecker symbol, and the action reads to order $n$:
\begin{eqnarray}
S[Q] = n S_{RS}[q]&=&\frac{1}{2}\left[G(\beta(1-q))+\beta q G'(\beta(1-q))-\beta^2 q(1-q)
G''(\beta(1-q))\right]\\ \nonumber
&+&\int_{-\infty}^{\infty}\,\frac{dz}{\sqrt{2\pi}}\,e^{-\frac{z^2}{2}}\ln\left[2 \cosh\left(\beta z \sqrt{q G''(\beta(1-q)}\right)\right] 
\end{eqnarray}
and the  derivative of  $S_{RS}[q]$ with respect to $q$ is
\begin{eqnarray}
\frac{dS_{RS}[q]}{dq}=\frac{\beta^2}{2}\left[G''(\beta(1-q))-\beta q
G'''(\beta(1-q))\right] \left[q-\int_{-\infty}^{\infty}\,\frac{dz}{\sqrt{2\pi}}\,e^{-\frac{z^2}{2}}\tanh^2\left(\beta z \sqrt{q G''(\beta(1-q)}\right)\right]
\end{eqnarray}
There are two replica symmetric saddle point equations 
\begin{equation}
G''(\beta(1-q))=\beta q G'''(\beta(1-q))
\end{equation}
and 
\begin{equation}\label{colqrs}
q=\int_{-\infty}^{\infty}\,\frac{dz}{\sqrt{2\pi}}\,e^{-\frac{z^2}{2}}\tanh^2\left(\beta z \sqrt{q G''(\beta(1-q)}\right)
\end{equation}
The first solution is unphysical \cite{deri} and thus $q$ is given by
Eq. (\ref{colqrs}). 
If we look for a second order phase transition,
expanding near $q=0$, we find that a continuous non-zero solution
can appear at $T_c=1/\beta_c$ where $\beta_c$ is determined by  
\begin{equation}\label{tcrs}
\beta_c^2 G''(\beta_c)=1
\end{equation}
This general equation was also derived in \cite{rom}. 
Now using Eq. (\ref{gprime}), Eq. (\ref{tcrs}) becomes
\begin{equation}\label{mucrs}
\int\,d\lambda \frac{\rho(\lambda)}{{(\mu_c-\lambda)}^2}=\infty
\end{equation}
and hence $\mu_c=\lambda_{\max}$ where $\lambda_{\max}$ is the 
largest eigenvalue of $J$. We thus find that $T_c$ is given by:
\begin{equation}\label{tccrs}
\frac{1}{T_c}=\int\,d\lambda \frac{\rho(\lambda)}{\lambda_{\max}-\lambda}
\end{equation}
It is straightforward to see that the possibility of having a finite
temperature phase transition in this Ising spin model depends on the 
existence of a finite temperature phase transition in the corresponding
spherical model. 
In this case the critical temperature  
$T_c$ of the two transitions are the same. 
The Eq. (\ref{tccrs}) determining $T_c$ shows that if
$\rho(\lambda)/(\lambda_{max}-\lambda)$ is integrable over  the
support of $\rho(\lambda)$ then $T_c$ is finite, if it is not integrable
then $T_c = 0$. Hence if $\rho(\lambda) \sim (\lambda_{\max} -\lambda)^\gamma$ 
near $\lambda_{max}$ then $T_c = 0$ for $\gamma \leq 0$ but a finite temperature
second order phase transition is possible for $\gamma > 0$. From 
Eq. (\ref{mucrs}) we also see that if $\gamma > 2$ then the phase transition 
can be continuous but of higher than second order. 

The above results can be further verified in a more general
than replica symmetric context by carrying out a Landau expansion. Writing
$Q_{ab}=\delta_{ab}+\omega_{ab}$, the lowest order expansion around
$\omega=0$ of Eq.(\ref{formgen}) is:
\begin{equation}\label{expand}
S[Q]=\frac{n}{2}G(\beta)+\frac{\beta^2}{2}G''(\beta)\left(\beta^2 G''(\beta)-1\right)\mbox{Tr}\,\omega^2+o(\omega^2)
\end{equation}
The nature of the phase transition depends on the coefficient
of $\mbox{Tr}\,\omega^2$ in the expansion above. This coefficient only vanishes
at $\beta^2 G''(\beta)=1$, which agrees with the previous definition of $T_c$.
If $T_c\neq 0$, then a second order phase transition occurs at
$T_c$, and the  subsequent replica symmetry breaking is determined by the
terms of higher order in $\omega$ in the expansion Eq. (\ref{expand}). 
We note that the breaking of the $O(n)$ symmetry in replica space should 
favor replica symmetry breaking \cite{parisi}. 

In addition if we look at the TAP
equations \cite{pp} in an external field, the linear expansion in the
paramagnetic phase gives the following equations for the magnetizations
$m_i$: 

\begin{equation}
m_i=\beta h_i+\beta \sum_{ij} J_{ij} m_j-\beta G'(\beta) m_i
\end{equation}
Hence the staggered susceptibility in the direction of an eigenvalue
$\lambda$ is

\begin{eqnarray}
\chi_{\lambda}&=&\frac{\beta}{1-\beta \lambda+\beta G'(\beta)}\\
&=&\frac{1}{\mu-\lambda}
\end{eqnarray}
from Eq.(\ref{gprime}). Hence, the staggered susceptibility for the maximum
eigenvalue $\lambda_{\max}$ diverges at the critical temperature, 
as in the SK model \cite{sk}.

We may now classify the phase transitions in the various models discussed
here simply by examining the behavior of 
the density of eigenvalues $\rho(\lambda)$ at its upper band edge.

\begin{itemize}
\item SK model (Eq. (\ref{dossk})): $\gamma = 1/2$  -- second order.
\item Hopfield $\beta >0$ (Eq. (\ref{doshl1}) and Eq. (\ref{doshg1}))): 
$\gamma = 1/2$ -- second order
\item Hopfield $\beta < 0$, $\alpha > 1$ (Eq. (\ref{doshg1})): 
$\gamma = 1/2$ -- second order.
\item Hopfield $\beta < 0$ , $\alpha < 1$  (Eq. (\ref{doshl1})): 
delta function at $\lambda_{\max}$
-- first order
\item SIMSK (Hopfield at $\alpha = 1$) $\beta < 0$  (Eq. (\ref{dossim})): 
$\gamma = -1/2$
-- first order.
\item ROM  (Eq. (\ref{dosrom})): delta function at $\lambda_{\max}$
-- first order.
\item Semi-square model: $\gamma = 0$ -- first order.
\end{itemize}
 
\subsection{One Step Replica Symmetry Breaking}
In a variety of models such as the SK model, the Hopfield model, or the
ROM, either the RS entropy or the annealed one is negative at low
temperature. In this case, replica symmetry has to be broken. Indeed, the
glass transition may be attributed to the existence of an extensive number
of pure states. The complexity of these pure states can be computed within
the following 1RSB ansatz \cite{kt,wemi}: $Q$ is a block diagonal
matrix, where the blocks have size $m \times m$ and $m \leq 1$. Inside
the blocks $Q_{ab}=(1-q)\delta_{ab}+q$. Then the action reduces to:

\begin{equation}
S[q,m]=\frac{m-1}{2m}G(\beta(1-q))+\frac{1}{2m}G(\beta(1-q+qm))-\frac{\lambda}{2}(1-q+mq)+\frac{1}{m}\ln\left(\int_{-\infty}^{\infty}\,\frac{dz}{\sqrt{2\pi}}\,e^{-\frac{z^2}{2}}\cosh^m(\sqrt{\lambda}z)\right)
\end{equation}
where:  
\begin{equation}
\lambda=\frac{\beta}{m}\left(G'(\beta(1-q+mq))-G'(\beta(1-q))\right)
\end{equation}
Expanding $S[q,m]$ around $m=1$ gives:
\begin{equation}
S[q,m]=-\beta f_{ann}+(m-1) V(q)+o\left((1-m)^2\right)
\end{equation}
The extremum of the effective potential $V(q)$ at $q=0$ contributes to the
paramagnetic value $-\beta f_{ann}$, whereas a local minimum at non zero $q$ 
corresponds to the entropy of pure states. The potential is easily computed:
\begin{eqnarray}
V(q)&=&\left.{\partial S[q,m]\over\partial m}\right|_{m=1}\\
&=&-\frac{1}{2}\left( G(\beta(1-q))-G(\beta)+\beta q G'(\beta)\right)+\frac{1+q}{2}\lambda-e^{-\frac{\lambda}{2}}\,\int_{-\infty}^{\infty}\,\frac{dz}{\sqrt{2\pi}}\,e^{-\frac{z^2}{2}}\cosh(\sqrt{\lambda}z)\ln\cosh(\sqrt{\lambda}z)
\end{eqnarray}
where:
\begin{equation}
\lambda=\beta\left(G'(\beta)-G'(\beta(1-q))\right)
\end{equation}
This is exactly the expression for the annealed complexity of the solutions
of the TAP equations found in \cite{pp}. 

The dynamical transition occurs when the number of pure states becomes
extensive. The dynamical transition temperature $T_D$ and the dynamical overlap
$q_D$ are determined from the equations: $V''(q_D)=V'(q_D)=0$ ($q_D\neq 0$). 
The static transition occurs when
the number of pure states is no longer extensive, so the static transition
temperature $T_K$ and the static overlap $q_S$ are determined from the
equations: $V(q_S)=V'(q_S)=0$ ($q_S\neq 0$). 

The first derivative of the potential is:
\begin{equation}
V'(q)=\frac{\beta^2}{2}G''(\beta(1-q))\left(q-\Gamma(q)\right)
\end{equation}
where
\begin{equation}
\Gamma(q)=e^{-\frac{\lambda}{2}}\,\int_{-\infty}^{\infty}\,\frac{dz}{\sqrt{2\pi}}\,e^{-\frac{z^2}{2}}\tanh^2(\sqrt{\lambda}z)\cosh(\sqrt{\lambda}z)
\end{equation}
and the second derivative is:
\begin{equation}
V''(q)=\frac{\beta^2}{2}G''(\beta(1-q))\left(1-\Gamma'(q)\right)
-\frac{\beta^3}{2}G'''(\beta(1-q))\left(q-\Gamma(q)\right)
\end{equation}
After some algebra, one finds for $q^*$ such that $V'(q^*)=0$:
\begin{equation}
V''(q^*)=\frac{\beta^2}{2}G''(\beta(1-q^*))\left(1-\beta^2 G''(\beta(1-q^*))\,
 e^{-\frac{\lambda}{2}}\,\int_{-\infty}^{\infty}\,\frac{dz}{\sqrt{2\pi}}\,e^{-\frac{z^2}{2}}\frac{1}{\cosh^3(\sqrt{\lambda}z)}\right)\label{v2}
\end{equation}
The paramagnetic solution $q=0$ is always a stationary point of $V(q)$ and
the second derivative of $V$ is:
\begin{equation}
V''(0)=\frac{\beta^2}{2} G''(\beta)\left(1-\beta^2 G''(\beta)\right)
\end{equation}

\begin{figure}[tbp]
\begin{center}
\rotatebox{0}
{\includegraphics*[width=9cm,height=9cm]{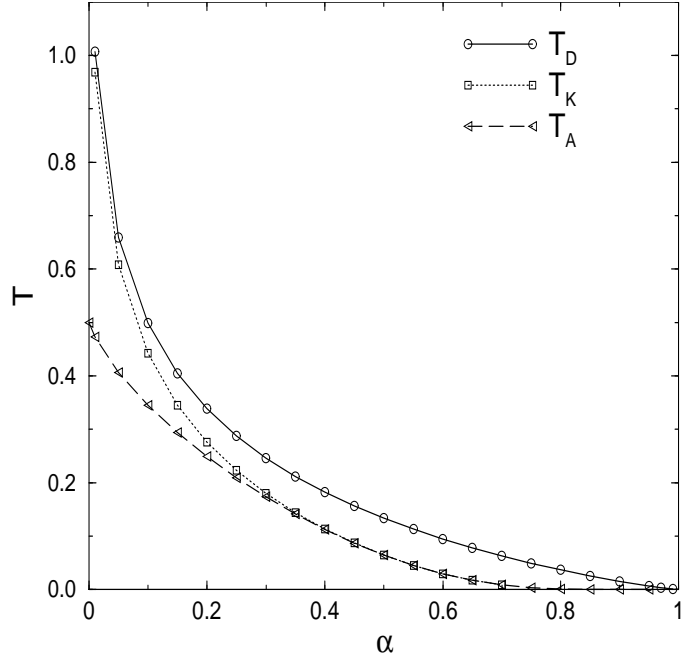}}
\caption{Various temperatures arising in the Random Orthogonal Model as a function of $\alpha$.}
\label{T.fig}
\end{center}
\end{figure}

Again, this quantity may have several possible behaviors, depending on $\beta_c$
\begin{itemize}
\item (i) $\beta_c=\infty$. One sees from	
Eq.(\ref{eqmu}) that for all temperatures $\beta^2 G''(\beta)<1$, and then 
$V''(0)>0$. Hence either the only local minimum of the effective potential
is at $q=0$, or there is another solution appearing at $T_D>0$ where the system
undergoes a dynamical transition, with a non zero dynamical overlap $q_D$.
\item (ii) $\beta_c=1/T_c<\infty$. From  Eq. (\ref{eqmu}), $\beta^2
G''(\beta)<1$ if $\beta<\beta_c$ and $\beta^2 G''(\beta)\geq 1$ if
$\beta\geq \beta_c$. Here, 
the stationary point $q=0$ becomes unstable at $T=T_c$ so there is no
dynamical transition below $T_c$ (one can not exclude a dynamical transition
at $T_D\neq0$, but as expected $T_c\leq T_D$). We remark from Eq. (\ref{v2})
that if $G'$ is convex, then there is no discontinuous dynamical transition 
and the system undergoes a classical spin glass transition at $T_c$.
\end{itemize}

This 1RSB calculation can be done for the Random Orthogonal Model, where a
discontinuous dynamical transition is expected. 
The dynamical and static temperatures $T_D$ and $T_K$ arising in the 
Random Orthogonal 
Model as a function of $\alpha$ are shown in Fig. (\ref{T.fig}). 
Also shown is the temperature $T_A$ at which the annealed entropy
disappears. It was noted by Marinari et al that for the ROM at
$\alpha=1/2$  $T_A$ is very close to $T_K$. Indeed we see that for all
$\alpha \geq 1/2$ this is the case. This means that the 
statics of these models for $\alpha > 1/2$ is very close to that of the
Random Energy Model (REM). This analogy is further supported by the 
values of $q$ in the one step solution which are already very close to $1$
at $T_K$.

\section{Numerical Simulations}
In this section we describe the numerical simulations carried out
to test our theoretical results. We shall concentrate on the 
case of the Random Orthogonal Model at different values of $\alpha$
and the semi-square model, which are  systems exhibiting the structural
glass transition. 

The numerical generation of the interaction matrices $J_{ij}$ is carried 
out as follows. We take a random orthonormal basis of ${\bf x}^{(k)}$ 
$1\leq k\leq N$ of  $\mathbb{R}^N$ and construct
$J_{ij}$ via
\begin{equation}
J_{ij} = \sum_{k} \lambda_k  x_i^{(k)}x_j^{(k)}
\end{equation}
where, in the case of a continuous density of eigenvalues $\rho(\lambda)$
 each $\lambda_k$ is drawn independently from the distribution with 
probability density $\rho(\lambda)$. In the case of the ROM,
in order to reduce sample to sample fluctuations, $\alpha N$
eigenvalues $+1$ and $N(1-\alpha)$ eigenvalues $-1$ are randomly assigned 
to each eigenvector.
As mentioned above, this has the form of a Hopfield model \cite{hop}
but where the patterns ${\bf x}^{(k)}$ are strictly orthogonal, 
and not simply statistically orthogonal, and where each pattern ${\bf
x}^{(k)}$ is weighted by $\lambda_k$.   
The construction of a statistically  $O(N)$ invariant basis ${\bf x}^{(k)}$
is carried out by choosing for the  ${\bf x}^{(k)}$ the (normalized) 
eigenvectors of the statistically $O(N)$ invariant symmetric Gaussian
matrix $K$ with $K_{ij} = \sigma_{ij}/\sqrt{N}$, each $\sigma_{ij}$ being 
Gaussian of mean $0$ and variance $1$.

\begin{figure}[tbp]
\begin{center}
\rotatebox{0}
{\includegraphics*[width=9cm]{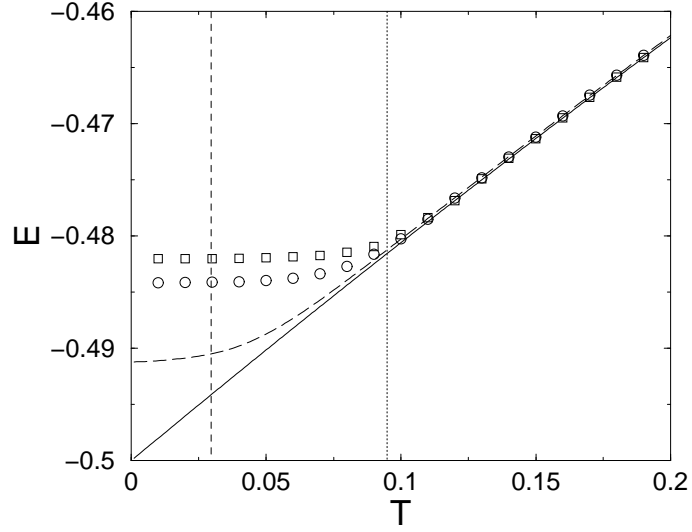}}
\caption{Energies per spin for the ROM with $\alpha=0.6$ : Monte Carlo 
simulations for systems of size $N=200$ (squares) and $N=30$ (circles), 
exact enumeration for $N=30$ (long dashed line), annealed (solid line). 
Also shown are the calculated values of $T_D$ (vertical dotted line) and 
$T_K$ (vertical dashed line).}\label{E06.fig}
\end{center}
\end{figure}

\begin{figure}[tbp]
\begin{center}
\rotatebox{0}
{\includegraphics*[width=9cm]{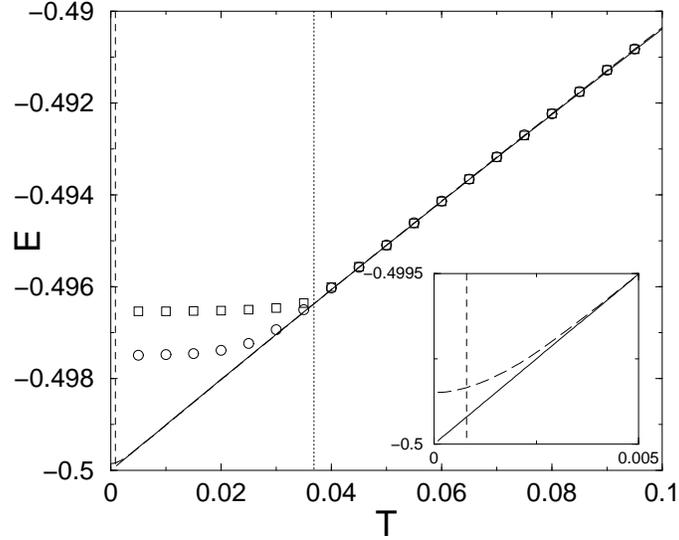}}
\caption{Energies per spin for the ROM with $\alpha=0.8$ : Monte Carlo 
simulations for systems of size $N=200$ (squares) and $N=30$ (circles), 
exact enumeration for $N=30$ (long dashed line), annealed (solid line). 
Also shown are the calculated values of $T_D$ (vertical dotted line) and 
$T_K$ (vertical dashed line). In the inset is shown the low energy behavior
of the annealed energy and the energy calculated by exact enumeration.}\label{E08.fig}
\end{center}
\end{figure}

\begin{figure}[tbp]
\begin{center}
\rotatebox{0}
{\includegraphics*[width=9cm]{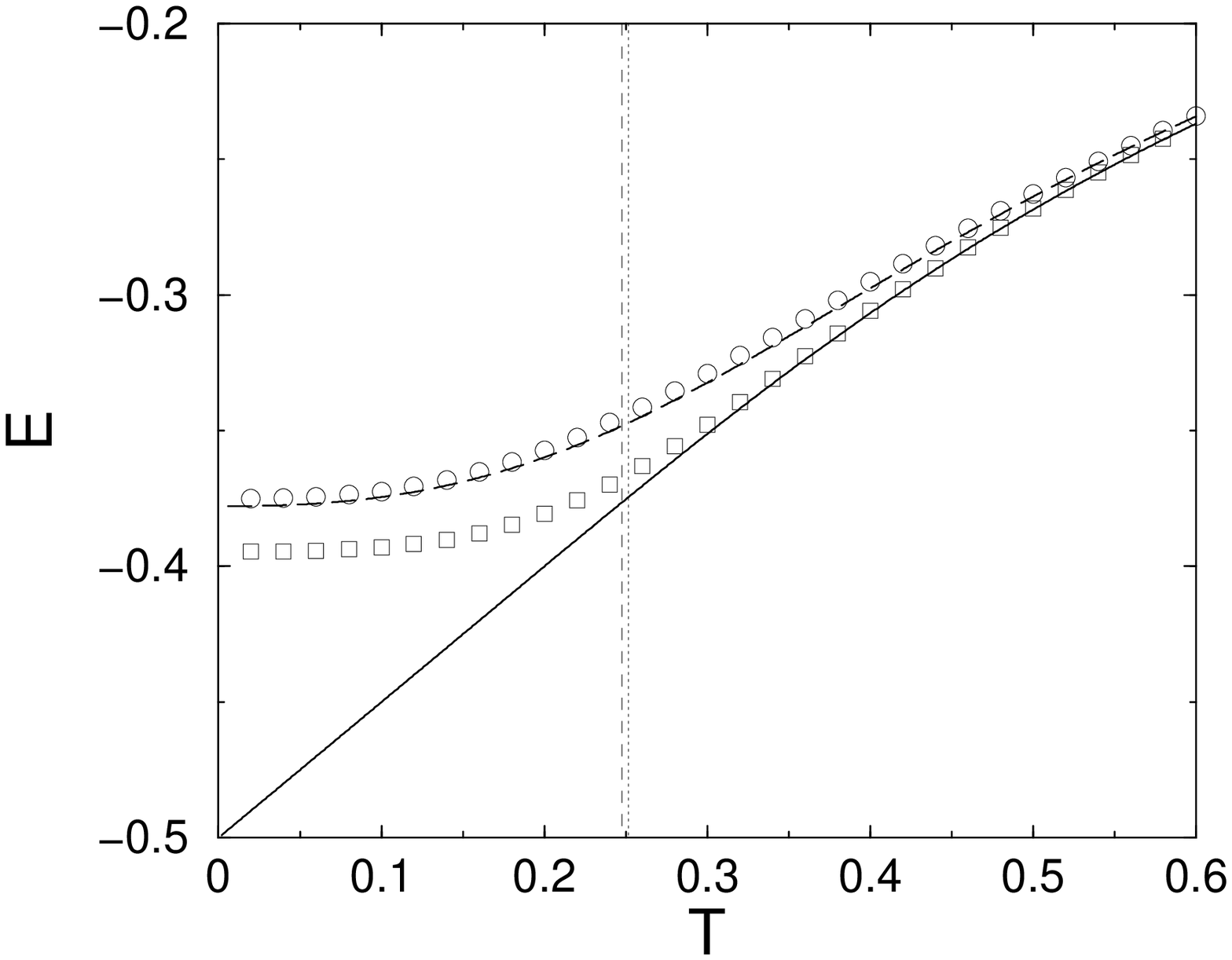}}
\end{center}
\caption{Energies per spin for the semi-square model : Monte Carlo 
simulations for systems size $N=200$ (squares) and $N=30$ (circles), 
exact enumeration for $N=30$ (long dashed line), annealed (solid line). 
Also shown are the calculated values of $T_D$ (vertical dotted line) and 
$T_K$ (vertical dashed line).}\label{EUnif.fig}
\end{figure}

We have carried out two types of numerical simulations. 
Monte Carlo simulations on systems of size $N=200$ were performed, 
in order to validate the high temperature predictions of the theory. Below
the dynamical transition temperature $T_D$ it is impracticable to 
equilibrate the system for these large system sizes, however 
one may estimate the value of $T_D$ by examining at which temperature
the measured results differ from the annealed calculation. The 
results of our calculations are compatible with these estimations.
In each of the Figs. (\ref{E06.fig}-\ref{E08.fig}-\ref{EUnif.fig}) is
shown the dynamically measured energy per spin for a system of size $200$. The
corresponding cooling rate, number of samples and number of runs are shown
table (\ref{tab1}). The equilibration time constituted $90$ percent of the time
spent at each temperature and the measurements were made during the last 
$10$ percent. Also shown on Figs.
(\ref{E06.fig}-\ref{E08.fig}-\ref{EUnif.fig}) is the calculated value of $T_D$
(vertical dotted line) and the value of $T_K$ (vertical dashed line). 
We see that for the system sizes studied here, the 
departure from the annealed energy and the onset of the 
characteristic, almost flat energy plateau, is in good agreement with 
the calculated value of $T_D$.

For system sizes of $N=30$ spins, 
the energy can be calculated by exact enumeration
over all the micro-states. The results for the energy can be compared with those
of dynamical simulations and the theoretical predictions. For the
dynamical simulations on systems of size $N=30$ the cooling rate, 
the number of samples and the number of runs are also indicated in table (\ref{tab1}).
The exact enumeration averages where taken over at least 20 samples. 
Also shown in Figs. (\ref{E06.fig}-\ref{E08.fig}-\ref{EUnif.fig}) 
are the results of these simulations. We see that the results of
the exact enumeration, even for the small system sizes used here, are in
excellent agreement with the theoretical predictions. For the system with
$\alpha = 0.6$ (Fig. (\ref{E06.fig})) we see 
that the plateau in the static energy is compatible
with the calculated value of $T_K$ but the Monte Carlo simulation with
$N=30$ is clearly out of equilibrium at temperatures below $T_D$. In 
Fig. (\ref{E08.fig}) for $\alpha = 0.8$ the theoretical prediction is
that $T_K \ll T_D$. We see that the exact enumeration result is
in perfect agreement with the annealed energy down to energies around $T_K$
(shown enlarged in the figure inset). The Monte Carlo results for systems
of size $N=30$ are however still clearly out of equilibrium. In 
Fig. (\ref{EUnif.fig}) are shown the results for the semi square model. 
We see that the dynamic and Kauzmann temperatures are
very close, however here the results of the Monte Carlo simulations
and exact enumeration for the systems of size $N=30$ are much closer, the 
dynamically measured energies are however still slightly lower
that the static ones measured by exact enumeration.   
\begin{table}
\begin{center}
\begin{tabular}[t]{|l|*{6}{c|}}
\hline
 & \multicolumn{2}{c|}{semi-square} & \multicolumn{2}{c|}{ROM $\alpha=0.6$} & \multicolumn{2}{c|}{ROM $\alpha=0.8$}\\
\cline{2-7}
 & {\small $N=30$} & {\small $N=200$} & {\small $N=30$} & {\small $N=200$} & {\small $N=30$} & {\small $N=200$}\\
\hline
number of samples & 40 & 40 & 40 & 40 & 40 & 40\\
\hline
number of runs & 200 & 20 & 200 & 20 & 200 & 20\\
\hline
cooling rate (MCS) & $5\times 10^5$ & $5\times 10^5$ & $10^6$ & $10^6$ &
 $2\times 10^6$ & $2\times 10^6$\\
\hline
\end{tabular}
\caption{Parameters used in the different Monte Carlo simulations.}\label{tab1}
\end{center}
\end{table}

\section{Conclusions}
In this paper we have examined the statics of a class of fully connected
generalized random orthogonal models. We have shown how the average over the 
$O(N)$ disorder can be carried out using a simple replica method recovering
the results of random matrix theory. This method has the useful property of
giving a variational form for the result.
Depending on the behavior of the 
density of eigenvalues at the band edges, we have seen that one either
obtains a classical spin glass transition or a structural glass transition. 
Our results suggest that in the thermodynamic limit only the density of
eigenvalues is important for the statics of these models. This classification 
should be useful in a wide range of models. It was noted that
the ROM  with a bimodal distribution of eigenvalues behaves like
a random energy model for large values of $\alpha$, in agreement with
previous studies where it was shown already to have very close
to REM-like behavior at $\alpha = 1/2$.

We have carried out numerical simulations on the generalized form of the 
original ROM model which support our analytical calculations. Simulating
small system sizes via Monte Carlo dynamics and by exact enumeration confirms 
the dynamical nature of the transition occurring at $T_D$. Further questions
arising form this study will be interesting to address. One can look at the
number of metastable states in such systems to better understand the 
geometric reasons leading to the glassy behavior \cite{cdl}. Also
the fact that even small system sizes stay out of equilibrium on
numerically accessible time scales and the fact that they can be studied 
by exact enumeration means that one may study finite size effects 
and hence activated processes on the dynamical transition as proposed 
in \cite{felix}. The formulation of the saddle point action in 
Eq. (\ref{actr}) also allows one to study the decomposition of the
Parisi overlap matrix $Q$ on the basis of eigenvectors of the problem,
this may give a more geometric picture of the nature of the glassy phase
of these models.

\end{document}